\documentclass[sigplan, screen, nonacm]{acmart} 
\acmConference[CC'25]{ACM SIGPLAN International Conference on Compiler Construction}{March 01--02, 2025}{Las Vegas, Nevada, United States}
\acmYear{2025}
\acmISBN{} % \acmISBN{978-x-xxxx-xxxx-x/YY/MM}
\acmDOI{10.1145/3708493.3712683} % \acmDOI{10.1145/nnnnnnn.nnnnnnn}
\startPage{1}

\setcopyright{none}
\usepackage{booktabs}   %% For formal tables:
\usepackage{subcaption} %% For complex figures with subfigures/subcaptions
\usepackage{microtype}
\usepackage{xcolor}
\definecolor{listinggreen}{rgb}{0,0.6,0}
\definecolor{listingkeywordcolor}{rgb}{1.0,0.4,0.0}
\definecolor{listinglightgray}{rgb}{0.8863,0.8863,0.8863}
\newcommand{\yes}{\color{listinggreen}{\textbf{\scriptsize \checkmark}}}
\newcommand{\no}{\color{listingkeywordcolor}{\textbf{\scriptsize $\times$}}}

\newcommand{\shrinkafter}{\vspace{-0.5cm}}

\begin{document}
\sloppy

%% Title information
\title[Data-efficient Performance Modeling via Pre-training]{Data-efficient Performance Modeling via Pre-training}         %% [Short Title] is optional;
                                        %% when present, will be used in
                                        %% header instead of Full Title.
%\titlenote{with title note}             %% \titlenote is optional;
                                        %% can be repeated if necessary;
                                        %% contents suppressed with 'anonymous'
%\subtitle{Subtitle}                     %% \subtitle is optional
%\subtitlenote{with subtitle note}       %% \subtitlenote is optional;
                                        %% can be repeated if necessary;
                                        %% contents suppressed with 'anonymous'

%% Author information
%% Contents and number of authors suppressed with 'anonymous'.
%% Each author should be introduced by \author, followed by
%% \authornote (optional), \orcid (optional), \affiliation, and
%% \email.
%% An author may have multiple affiliations and/or emails; repeat the
%% appropriate command.
%% Many elements are not rendered, but should be provided for metadata
%% extraction tools.

%% Author with single affiliation.
\author{Chunting Liu}
%\authornote{with author1 note}          %% \authornote is optional;
                                        %% can be repeated if necessary
%\orcid{nnnn-nnnn-nnnn-nnnn}             %% \orcid is optional
\affiliation{
  %\position{Position1}
  %\department{Department1}              %% \department is recommended
  \institution{New York University Abu Dhabi}     %% \institution is required
  %\streetaddress{Street1 Address1}
  \city{Abu Dhabi}
  %\state{State1}
  %\postcode{Post-Code1}
  \country{UAE}                    %% \country is recommended
}
\email{cl5503@nyu.edu}          %% \email is recommended

\author{Riyadh Baghdadi}
%\authornote{with author1 note}          %% \authornote is optional;
                                        %% can be repeated if necessary
%\orcid{nnnn-nnnn-nnnn-nnnn}             %% \orcid is optional
\affiliation{
  %\position{Position1}
  %\department{Department1}              %% \department is recommended
  \institution{New York University Abu Dhabi}     %% \institution is required
  %\streetaddress{Street1 Address1}
  \city{Abu Dhabi}
  %\state{State1}
  %\postcode{Post-Code1}
  \country{UAE}                    %% \country is recommended
}
\email{baghdadi@nyu.edu}          %% \email is recommended

%% Abstract
%% Note: \begin{abstract}...\end{abstract} environment must come
%% before \maketitle command
\begin{abstract}
Performance models are essential for automatic code optimization, enabling compilers to predict the effects of code transformations on performance and guide search for optimal transformations. Building state-of-the-art performance models with deep learning, however, requires vast labeled datasets of random programs -- an expensive and time-consuming process, stretching over months. This paper introduces a self-supervised pre-training scheme with autoencoders to reduce the need for labeled data. By pre-training on a large dataset of random programs, the autoencoder learns representations of code and transformations, which are then used to embed programs for the performance model. Implemented in the Tiramisu autoscheduler, our approach improves model accuracy with less data. For example, to achieve a MAPE of $20.72\%$, the original model requires 18 million data points, whereas our method achieves a similar MAPE of $22.44\%$ with only 3.6 million data points, reducing data requirements by $5\times$.

\end{abstract}

%% 2012 ACM Computing Classification System (CSS) concepts
%% Generate at 'http://dl.acm.org/ccs/ccs.cfm'.

%% End of generated code

%% Keywords
%% comma separated list
  %% \keywords are mandatory in final camera-ready submission

%% \maketitle
%% Note: \maketitle command must come after title commands, author
%% commands, abstract environment, Computing Classification System
%% environment and commands, and keywords command.
\maketitle

\section{Introduction}
State-of-the-art compilers have made significant progress in accelerating compute-intensive applications such as deep learning, image processing, and scientific computing. This is done thanks to the application of complex program and data layout transformations, such as loop fission, fusion, parallelization, and vectorization~\cite{baghdadi_pencil_specification, Baghdadi2019, Ragan2013}. Many state-of-the-art compilers~\cite{Chen2018, Zheng_2020, Adams2019, Kaufman2021ALP, Baghdadi2021ADL, Merouani_2024} heavily rely on performance models to guide their decision-making. Such an approach of automatic code optimization involves using a search technique to explore the space of possible code transformations, selecting candidates, and finally, evaluating and choosing the candidate that minimizes execution time. In this context, performance models are used to evaluate transformations without running the code during compilation, which results in a much faster compilation. Significant research has focused on developing performance models with high accuracy. In particular, recent work has employed deep-learning-based performance models to address the complexity of the problem and provide accurate evaluations~\cite{mendis2019ithemal, Adams2019, Chen2018, Baghdadi2021ADL, Merouani_2024}.

Building performance models is challenging since it requires generating a large dataset of random programs (millions of data points). Moreover, labeling such an amount of data is expensive computationally (stretching over months). This is because for each randomly generated program, code optimizations are sampled from the search space; then the code is compiled and run multiple times (to obtain a stable measurement). Each run might take a few seconds up to a few hours. Repeating this process millions of times is extremely time-consuming. For example, the DNN-based performance model used in Tiramisu \cite{Merouani_2024} was trained on a dataset of $26$ million datapoints, which took $6$ months to generate on a 15-node multicore CPU cluster. These large computational requirements to generate the training data limit the development and practical use of DNN-based performance models.

This demand for large amounts of data can be partly attributed to the difficulty in learning an encoding for the input programs and code transformations to be applied. Let us take the example of the performance model used in Tiramisu~\cite{Baghdadi2021ADL}. It encodes programs by extracting a set of simple features and concatenating them into vectors. This representation, although easy to extract, lacks sufficient abstraction for effective learning~\cite{Bengio_2013}. The model has to learn how to combine the simple features into complex and comprehensive ones, demanding a large dataset that is expensive to generate. While some other DNN-based performance models, such as Halide's model~\cite{Adams2019}, require less data since they directly extract complex hand-engineered features from the input code (more than 57 complex hand-engineered features). Extracting such features is complex, which adds a significant burden on the compiler developers and is error-prone. In this work, we focus on the class of DNN-based performance models that take simple features as input. Such features are easy to extract from source code, which reduces the burden on the compiler developers and the possibility of bugs in feature extraction.

To address the expensive data requirements in training DNN-based performance models, we drew inspiration from pre-training techniques widely employed in domains such as computer vision \cite{Tong_2022, Zhang_2022, Ding_Yan_Wang_Zhao_Zhuang_Xu_Li_2023} and natural language processing \cite{Lugosch_Ravanelli_Ignoto_Tomar_Bengio_2019, Montero_2021, Freitag_2018}. Pre-training allows a model to learn general and meaningful features from large datasets. Once the pre-training is done, it can be used to extract effective embeddings from the input, hence reducing the data requirement of the new models that use such embeddings as input.   

In this work, we propose a pre-training method that uses an autoencoder to learn the representation of programs. The autoencoder is trained to encode and reconstruct program statements, and this is done in an unsupervised way so that the expensive data labeling is avoided. The encoder part of the pre-trained autoencoder is then used to embed program statements before feeding them into the performance model. This reduces the data required to train the performance models as an effective embedding of the statements is already learned by the encoder. While some pre-training methods for code have been proposed in the literature, such methods are not suitable for the problem of automatic code optimization. Some of them require code compilation \cite{Ben_2018, VenkataKeerthy_2020}, which significantly increase the search time (discussed in Sec. \ref{subsection:design-choices}). While others are designed to only model code representation but not code optimizations \cite{kanade_2020_cubert, guo_2022_unixcoder, guo_2020_graphcodebert, cummins_2023}. Our proposed approach is the first to be demonstrated in the context of code optimization.

We implemented the proposed approach in the Tiramisu performance model~\cite{Merouani_2024}, a state-of-the-art performance model. We choose Tiramisu's performance model as the baseline because it extracts high-level features directly from the source code, bypassing the costly compilation process when exploring possible code optimizations. Our evaluation demonstrates that the proposed approach significantly improves the accuracy of the Tiramisu performance model when the training dataset is small.
For example, in order to achieve a MAPE (Mean Absolute Percentage Error) of $20.51\%$, the original Tiramisu performance model requires 18 million data points, whereas it only requires $3.6$ million data points to reach a comparable MAPE of $22.44\%$ when our pre-training approach is used, reducing the amount of data needed by $5\times$.
Surprisingly, we found that even when training with $40\times$ less data ($0.45$ million data points), the Tiramisu performance model achieved a MAPE of only $29.69\%$ when our pre-training approach was used, in contrast to $37.27\%$ when our approach was not.

The contributions of this paper are as follows:
\begin{itemize}
    \item We propose a pre-training method based on auto-encoders for DNN-based performance models. 
    \item We implement and evaluate the proposed method and demonstrate its effectiveness in reducing the data requirements for training performance models.
    \item We release the proposed pre-trained model, along with the pre-training dataset to the scientific community \footnote{\url{https://github.com/Tiramisu-Compiler/cost_model_pretrain}}. %\footnote{\url{https://github.com/Tiramisu-Compiler/cost_model_pretrain}}.
\end{itemize}

% Move to after the introduction
% Add a table to summarize differences on several dimensions: 
% Features level 
% Compilation 
% Speedup prediction (yes/no)
% System profiling (pretraining task: easy/complex)
% 
\section{Related Work}

In this section, we provide an overview of compilers that use a search-based method and a learned performance model for automatic code optimization. We also present work that uses pre-training to learn code embeddings. Table~\ref{tab:related} shows a summarized comparison between state-of-the-art pre-training methods. Finally, we present existing work that addresses the problem of reducing the data requirements for training machine learning models used within compilers.

\begin{table}[tb]
    \scriptsize
    \vspace{0.5cm}
    \setlength\tabcolsep{2pt}
    \caption{\textsc{Qualititive Comparison with related works.}}
    \begin{tabular}{|p{4cm}|c|c|c|c|c|c| }
        \hline
        
        \textbf{Features} & \textbf{\rotatebox{90}{\ Ours\ }} & \textbf{\rotatebox{90}{\ Inst2Vec~\cite{Ben_2018}\ }} & \textbf{\rotatebox{90}{\ IR2Vec~\cite{VenkataKeerthy_2020}\ }} & \textbf{\rotatebox{90}{\ Tr\"{u}mper et al.~\cite{Trumper_2023}\ }} & \textbf{\rotatebox{90}{\ Selvam et al.~\cite{selvam_2023}\ }} & \textbf{\rotatebox{90}{\ Sasaki et al.~\cite{Sasaki_2022}\ }}  \\\hline

        \textbf{Operates on High-level IR}         & \yes  & \no  & \no   & \yes & \yes & \yes \\\hline

        \textbf{Supports High-level Optimizations} & \yes & \no   & \no   & \yes & \yes & \yes \\\hline
        
        \textbf{Does Not Require Compilation}       & \yes  & \no & \no   & \yes & \yes & \yes \\\hline

        \textbf{Does Not Require Labeling}          & \yes  & \yes & \yes & \no & \yes & \no \\\hline

        \textbf{Evaluated on Performance Modeling} & \yes  & \no   & \no  & \yes & \yes & \yes \\\hline

        \textbf{Supports General Loop Nests}        & \yes  & \yes & \yes & \yes & \no & \yes \\\hline

        \textbf{Architecture Independent}           & \yes  & \yes & \yes & \no & \yes & \no \\\hline

    \end{tabular}
    \label{tab:related}
\end{table}

\subsection{Search-based Compilers with Learned Performance Models}

Many compilers use a search-based method with a learned performance model. Examples include TVM~\cite{Chen2018, Zheng_2020}, Halide~\cite{Adams2019}, Tiramisu~\cite{Baghdadi2021ADL}, and XLA~\cite{Kaufman2021ALP}. These compilers take high-level code or computation graphs as input and employ search algorithms such as Monte Carlo tree search (MCTS)~\cite{Baghdadi2021ADL}, Genetic Algorithm (GA)~\cite{Zheng_2020}, Beam Search~\cite{Adams2019}, Simulated Annealing~\cite{Chen2018}, and Reinforcement Learning~\cite{Byung_2020} to explore combinations of high-level code optimizations such as loop tiling, vectorization, parallelization, unrolling, and fusion.
These search-based methods usually have two components: a search space exploration module and an evaluation module. The role of the search space exploration module is to explore the space of code optimizations that optimize a given program. The evaluation module is in charge of assessing the quality of candidates that are encountered during the exploration. This module consists of a deep learning performance model that is trained to predict the potential quality (e.g., execution time or speedup) that a sequence of code optimizations would yield if it was applied to the input program. In this work, we focus on proposing a pre-training method for the performance model used in one of these compilers, Tiramisu~\cite{Baghdadi2021ADL}, but our proposed method can, in principle, be adapted to other compilers similar to Tiramisu, as long as their performance models take simple features as input (since the goal of pre-training is to learn a rich embedding from the simple features extracted from code).

\subsection{Pre-training to Learn a Code Representation}
\label{subsection:learn-code}

Our proposed pre-training approach is similar to the idea of transfer learning: adapting a trained model to a new but similar task. In our case, the encoder learns an effective code representation, and this representation is then used to train for speedup prediction. Previous work has explored the use of pre-training to learn code representations, which can then be used to perform various tasks. There are two levels of code from which features are commonly extracted: source-level code and low-level IR (Intermediate Representation), for example, the LLVM IR. Typical source-level code features used for pre-training include code token sequences~\cite{kanade_2020_cubert}, abstract syntax trees~\cite{guo_2022_unixcoder}, data flow graphs~\cite{guo_2020_graphcodebert}, etc. Cummins et al.~\cite{cummins_2023} even utilize large language models to learn code representations. However, these models that rely on source-level code features for pre-training focus on tasks such as code search, code classification, and code generation. Our proposed approach also pre-trains on source-level code features, but is designed for the task of speedup prediction.

Work such as \texttt{Inst2Vec}~\cite{Ben_2018} and IR2Vec~\cite{VenkataKeerthy_2020} learn embeddings from a low-level IR (LLVM IR). The learned embeddings are then fed to deep learning models for a variety of tasks such as algorithm classification, mapping to heterogeneous devices, and predicting the best thread coarsening factor. However, using LLVM IR-based embeddings in a search-based compiler is costly. Therefore, they are not suitable for search-based compilers considered in this paper. The main issue is that autoschedulers that use a search-based method explore a large space of code optimizations. They then use the performance model to evaluate the quality of each candidate they visit in the space. In order to extract a representation from the LLVM IR level, code needs to be compiled down to LLVM IR first, which is time-consuming when done millions of times.
As an example, the Halide autoscheduler~\cite{Adams2019} evaluates millions of candidates in the search space. For a performance model to be well suited for search-based autoschedulers, it should ideally predict performance from the source-level directly without the need for compilation. We provide more details about this issue later in Sec. \ref{subsection:design-choices}.

Tr\"{u}mper et al. \cite{Trumper_2023} propose a similarity-based approach that allows the knowledge from pre-trained embeddings to be transferred between similar loop nests. Their pre-training requires predicting system-specific metrics such as the main/L3/L2 memory bandwidth and data locality, and thus the embeddings they learn are system-specific. They are mainly used to train new tasks for the same machine. This is unlike our embeddings, which are machine-independent. Our embeddings are learned by encoding and decoding code and do not use system-specific information, and therefore the same embeddings could be used in multiple performance models, each targeting a different hardware, which simplifies the development of performance models. In addition, collecting system-specific metrics such as the memory bandwidth requires code execution, which is time-consuming. Our goal in this paper is to develop a pre-training method that does not require code execution, since code execution is time-consuming.

\citet{selvam_2023} use a graph autoencoder to learn representations of unlabeled deep learning graphs, then combine it with a supervised graph neural network training to predict metrics such as memory usage and step time. Unlike our proposed method, this work is domain-specific. It is mainly designed to learn embeddings from deep learning graphs. Our work is more general. First, it learns embedding from source code that has loops, arrays, statements, etc. Second, it supports multiple domains, including deep learning, image processing, linear algebra, stencils, etc.

Unlike all of the previously mentioned projects, our work has the uniqueness of being trained and evaluated on the task of speedup prediction. We believe that speedup prediction, in particular, is a hard task due to the complexity of the underlying hardware, and the intricate relationship between code and code optimizations and also among code optimizations themselves.

Sasaki et al. \cite{Sasaki_2022} also utilize pre-training techniques to alleviate the high data requirements of performance modeling. Their approach allows a user to train a performance model for a given target machine and then port the model to a new machine by fine-tuning the model on a small amount of data generated on the new machine. The major difference between their approach and ours is that our pre-training step does not require the expensive data labeling that they do. In their case, the user needs to use an initial large dataset, collected on a given hardware, as a pre-training method. We believe that requiring the user to collect such a large dataset hinders the development of performance models. In our case, the user can still benefit from pre-training even if they do not have such a large labeled dataset.

\subsection{Reducing Data Requirements}

\citet{leather_2009} and \citet{ogilvie_2017} also have the objective of reducing the high cost of program profiling when generating data for training. Our approach is complementary to theirs. Their primary goal is to minimize the number of optimizations that need to be explored for each program in their dataset (number of optimizations per program), whereas we aim to minimize the total number of data points $(\text{number of programs} \times \text{number of optimizations per program})$ required to train a DNN-based performance model. An interesting direction for future research could be to apply our approach in conjunction with theirs to further reduce the data requirements.
%for training an accurate DNN-based performance model. 
%Their primary distinction between their work and ours is that they focus on minimizing the number of optimizations that need to be explored for a single program, whereas we aim to minimize the total number of data points $(\text{number\_of\_programs} \times \text{number\_of\_optimizations\_per\_program} )$ required to train a DNN-based performance model. An interesting direction for future research could be to apply our approach in conjunction with theirs to further reduce the data requirements for training an accurate DNN-based performance model. 

\section{Background}
In this paper, we use Tiramisu's performance model as a representative models that only relies on features that can be extracted from source code. This section provides an overview of Tiramisu's autoscheduler, and the DNN-based performance model that it uses.

\subsection{Autoscheduling in Tiramisu}
Tiramisu \cite{Baghdadi2019} is a polyhedral compiler that uses a deep-learning-based performance model \cite{Baghdadi2021ADL} to explore code transformations. The polyhedral model serves as a comprehensive mathematical framework for the representation of code and code transformations, facilitating reasoning about the correctness of transformations \cite{pencil_2015, paul_2011, Sven_2010}. This model extracts information such as the iteration domain, access relations, and schedule for each code statement. Different code transformations are implemented by modifying the schedule, which changes the order of execution of statement instances in the iteration domain.

To automatically pick the best sequence of transformations, Tiramisu's autoscheduler performs a tree search to explore the space of valid transformations. The root represents the unoptimized code and each of the other nodes represents one particular transformation. The path from the root to a particular node is then a sequence of transformations. The search tree is expanded level by level, and the performance model is used to evaluate which branches of transformations yield the highest speedups and should be further explored. Exploration stops after reaching a pre-defined search depth $L$, and the performance model is responsible for evaluating and picking the transformation sequence that yields the best speedup. 

\subsection{Performance Modeling using Deep Learning in Tiramisu}
The DNN (Deep Neural Network) based performance model developed by Merouani et al. \cite{Merouani_2024}, an updated model from \cite{Baghdadi2021ADL}, supports programs that can be expressed in Tiramisu. The objective of the performance model is to predict the speedup of a given code when a sequence of code transformations is applied on it. Since hand-engineering features for speedup prediction is a tedious task, the performance model proposed by \citet{Merouani_2024} extracts simple high-level information about the program and stores them as ordered, variable-sized set of compact vectors, called computation vectors. Each computation vector corresponds to a statement. The performance model recursively embeds these vectors based on the AST (Abstract Syntax Tree) representation of the program, and the final embedding is fed to a fully-connected network to predict speedup. 

\subsubsection{Input representation} \label{subsection: input_representation}
The input of the performance model is the unoptimized code and the optimization sequence that is to be applied on it. \citet{Merouani_2024} extract features from the program statements to form computation vectors. The performance model then organizes these computation vectors as the leaves of the code's AST, with the other nodes representing information about each loop level. Since Tiramisu is a polyhedral compiler, many of the features that it extracts are a part of the polyhedral representation of the code. Computation vector encodes the following information about statements in the program: 

\begin{itemize}
    \item \textbf{Iteration domain matrix}: A matrix that represents the iteration domain of a statement (polyhedral representation), which refers to the range or set of values the iterators of all the loops containing this statement. 
    \item \textbf{List of access matrices}: In the polyhedral model, an access to a memory buffer is represented as an access matrix \cite{paul_2011}. The access matrix has $k$ rows and $n + 1$ columns, where $k$ is the number of dimensions of the access buffer and $n$ is the loop depth. Each row in the matrix represents an array dimension. Each array dimension is considered to be a linear combination of the loop iterators. The coefficient of each loop iterator is stored in the column that corresponds to that loop iterator. The last column in the matrix corresponds to constants. For example, the memory access $A[i_0, i_0+i_1, i_1-2]$ can be represented by the matrix $M$: 
    $$M = \begin{bmatrix}
            1 & 0 & 0\\
            1 & 1 & 0 \\
            0 & 1 & -2
            \end{bmatrix}.$$ 
    Each row of $M$ corresponds to each of the access dimension. We see that the third dimension of the access is $i_1-2$, so the third row has entries $0, 1, -2$, which can be also written as $0\times i_0+1\times i_1 -2$.
    \item \textbf{Operations vectors}: Each operation $(+, -, \times, \div, etc.)$ on the right hand side of the assignment is encoded as a one-hot vector. The vectors representing the operations of the same statement are then concatenated together with a post-order traversal of the expression tree.

    \item \textbf{Schedule matrices}: A sequence of transformation matrices that encodes the sequence of affine transformations (the polyhedral schedule matrix representation is used). The supported transformations that are represented as matrices include loop skewing, reversal, interchange, fusion, and distribution.

    \item \textbf{Transformation features}: It encodes the other  transformations that are not encoded in the schedule matrix, and which include parallelization, tiling and unrolling.
    
\end{itemize}

\subsubsection{Model Architecture}
\begin{figure}
    \centering
    \includegraphics[width = 0.17\textwidth]{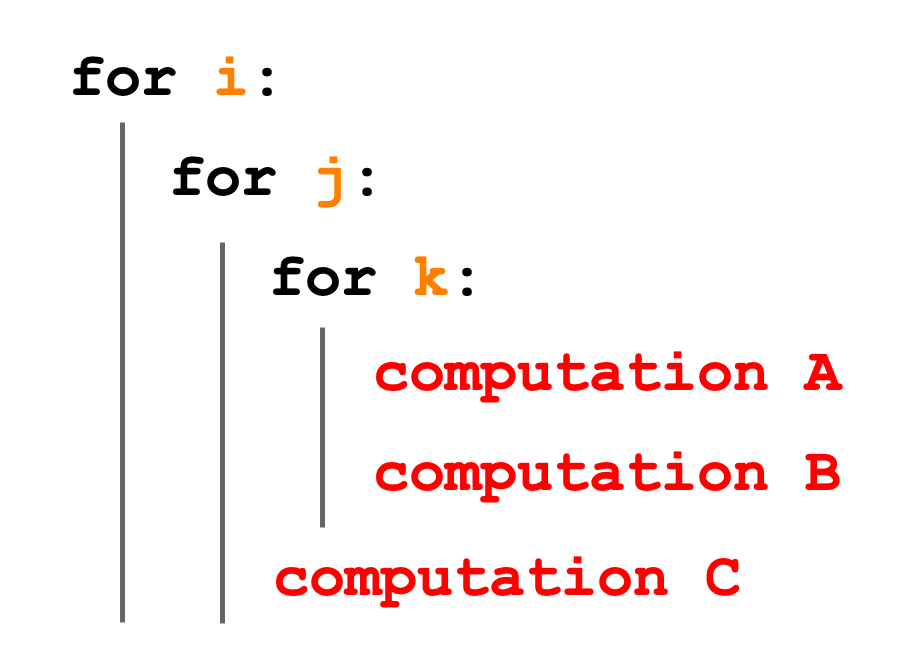}
    \caption{An example nested for loop structure.}
    \label{for_loops_structure}
\end{figure}

\begin{figure}
    \centering
         \includegraphics[width = 0.48\textwidth]{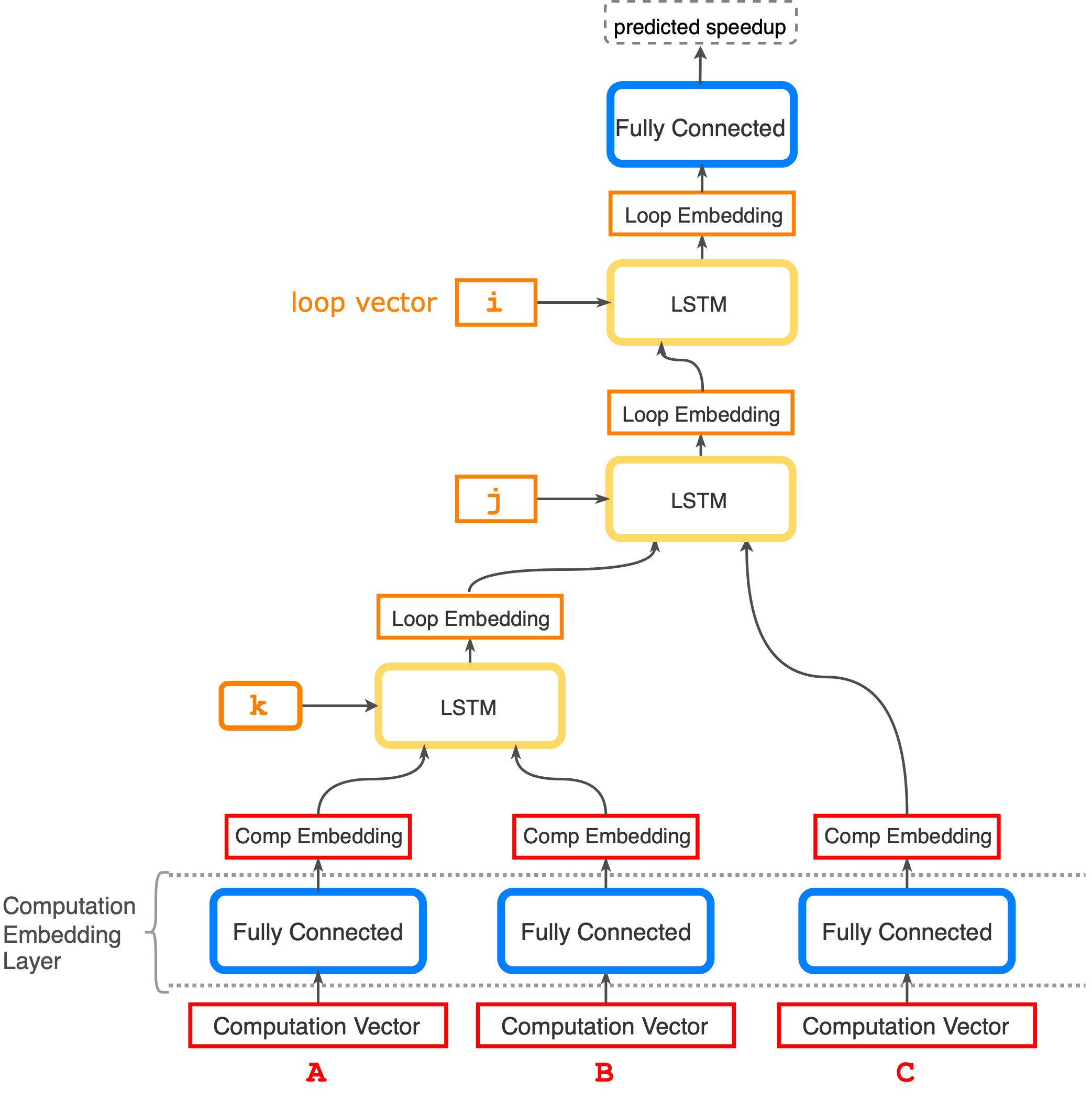}
         \caption{DNN based performance model by \citet{Merouani_2024}}
         \label{fig:performance_model_dnn}
         \shrinkafter
\end{figure}

The architecture of the DNN-based performance model by \citet{Merouani_2024} is dynamically structured according to the AST of the input program. For example, one can map the nested loop program in Figure \ref{for_loops_structure} to the DNN architecture in Figure \ref{fig:performance_model_dnn} based on its AST. Although the AST structure differs according to the input program, they all consist of the following basic components: (1) A fully connected network that embeds each computation vector into a computation embedding, (2) an LSTM network that embeds all the child computation vectors and loop embedding vectors into a loop embedding vector, and (3) a fully connected network that takes the final loop embedding vectors to predict the speedup.

\section{Data Requirement Challenge}
While the deep learning-based performance model by~\citet{Merouani_2024} highlighted previously has shown general applicability across a diverse range of code transformations and demonstrates high accuracy in speedup prediction, it was trained on a large dataset, containing 26 million data points. Generating this dataset demands substantial time and resources, as measuring the running time of programs in the dataset implies  compiling and running them. Moreover, for a given data point, multiple runs are required to reduce the effects of noise on measurement. Generating this dataset took $6$ months on a 15-node multicore CPU cluster. This extensive time and resource consumption during data generation makes the development of similar models difficult.

\begin{figure}
    \centering
    \includegraphics[width=0.48\textwidth]{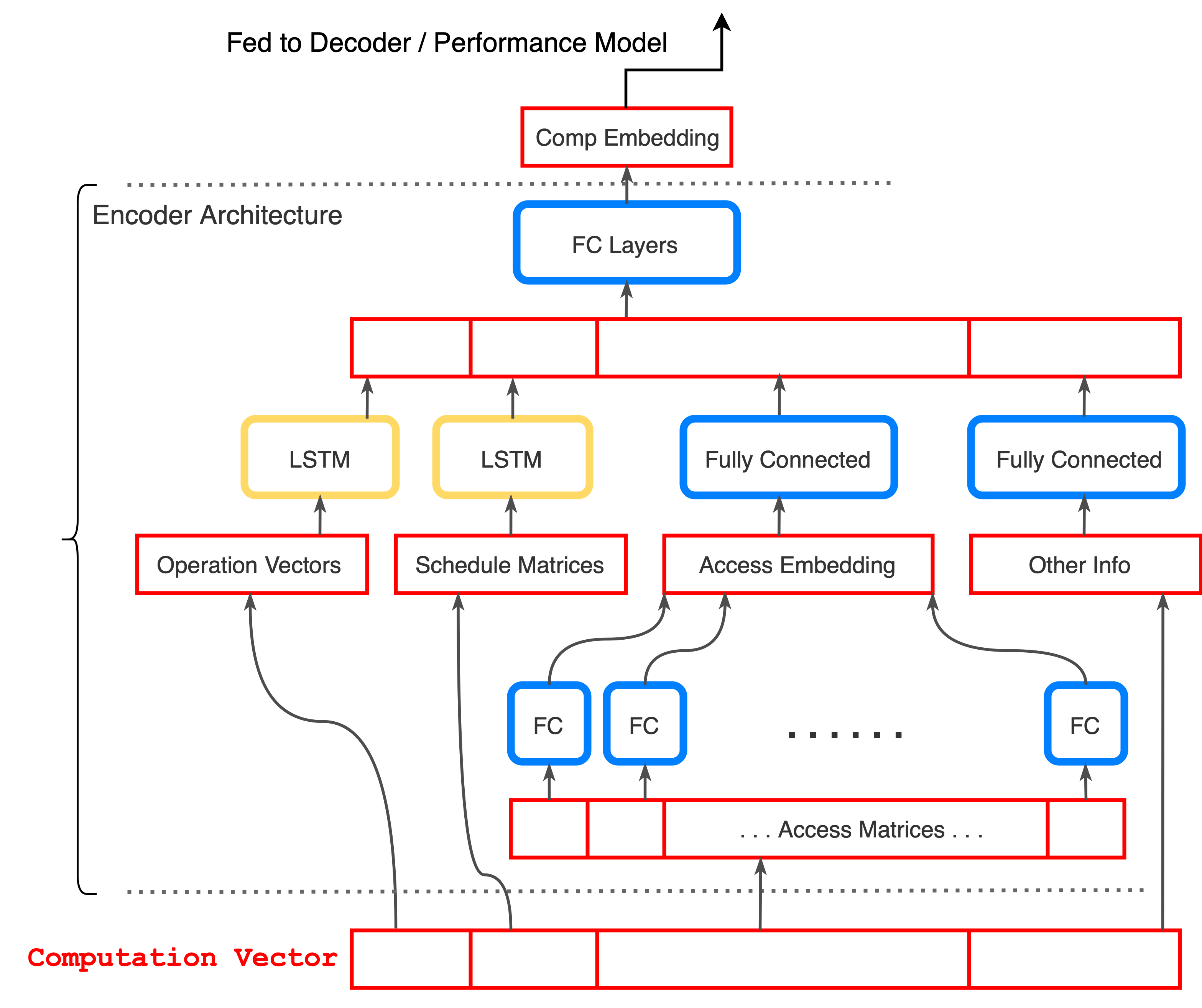}
    \caption{Architecture of the Encoder Part of the Autoencoder}
    \label{fig:complex_encoder}
    \shrinkafter
\end{figure}

\section{Autoencoder-based Pre-training} \label{methodology}
To address the data requirement challenge, we propose to use an autoencoder-based pre-training. The motivation behind this is that the Tiramisu performance model encodes programs by extracting simple features encoded in vectors, which can lack sufficient abstraction for effective learning \cite{Bengio_2013}. The rationale behind using pre-training is the following: the original model has to learn two tasks simultaneously: 1) embeddings of programs and optimizations; and 2) how to map these embeddings to speedups. Learning these two, using labeled data, is likely to require more data compared to learning just the second (mapping embeddings to speedups). By using a pre-trained model, that has learned high quality embeddings of programs and optimizations using self-supervised learning, the model will require less labeled data to map those embeddings to speedups. 

\subsection{Workflow Overview}

We first pre-train an autoencoder to embed computation vectors. We use a large dataset of computation vectors to do this (a computation vector is a vector that represents a statement; the composition of such a vector is discussed in Sec.~\ref{subsection: input_representation}). This unsupervised learning technique avoids the need for expensive code execution that performance models suffer from. Once the autoencoder is trained, its encoder part is used as a pre-processing step to embed computation vectors before they are fed to train the performance model. The weights in the encoder network are frozen at the beginning of the training. After the loss of the performance model stabilizes, we allow the weights of the encoder to be updated by backpropagation. This final step is also called fine-tuning. The rest of this section provides more details about each of these steps.

\subsection{Pre-training Data}
\label{subsection:pre-training-data}
We use the same code generator used in LOOPer~\cite{Merouani_2024} to generate a large number of random programs. For each randomly generated program, candidate code transformations are sampled for the space of possible code transformations, using LOOPer's search technique. This process generates pairs of Tiramisu programs and the code transformation sequences that could be applied to them.
For each statement in the previously generated data, we create a computation vector that has the same composition as discussed in Sec.~\ref{subsection: input_representation} (the computation vector in this case is a set of simple features representing the statement, its iteration domain, and the code transformations applied to it). Note that this process does not require compiling and running programs, making the generation of this pre-training dataset cheap computationally. The dataset of computation vectors we extracted to train our autoencoder consists of $26$ million computation vectors. Each computation vector represents a datapoint in the pre-training dataset.

\subsection{Autoencoder-based Pre-training} \label{subsection:autoencoder-pretraining}
As discussed in the previous section, the input of the DNN-based performance model is the statements of the program encoded as computation vectors. If we learn an embedding of computation vectors (which consist of complex features such as schedules, iteration domains, and memory accesses), the performance model may require less data when it is trained to make speedup prediction. To achieve this, we carefully devised an encoder architecture (depicted in Figure \ref{fig:complex_encoder}) to embed computation vectors, with a simpler decoder comprising multiple layers of fully connected networks. Unlike the computation embedding layer in Figure \ref{fig:performance_model_dnn}, which processes the entire computation vector with a deep fully connected network, our proposed encoder dissects the computation vector and embeds each component separately. For example, every access matrix in the computation vectors is fed to a fully connected (FC) network. The outputs of these FC networks are concatenated to form an embedding for the access matrices. All the component embeddings are then concatenated and passed through layers of FC to generate computation embeddings. In the pre-training phase, these embeddings are fed to the decoder to reconstruct the input computation vector, the Mean Square Error (MSE) of the reconstruction serving as the loss function. Note that the encoder is a deeper and more complex network compared to the computation embedding layers in the original performance model proposed by \citet{Merouani_2024}. This deeper network can potentially increase the model's ability to extract more abstract and meaningful features from the input statements, which is useful for performance modeling.

Training an autoencoder as a pre-training task provides many benefits. Due to its unsupervised nature, it circumvents the need for expensive speedup measurements. Moreover, the embedding is learned rather than hand-engineered. Our team has explored ways of encoding the input program using simple hand-engineered features for speedup prediction, and none of them have shown effectiveness in alleviating the data requirement problem. Hand-engineering features is hard since one needs to know precisely which features to use, without missing any important one. In addition, feature extraction has to be implemented in the compiler, which adds burden on the compiler developers. Any bugs in feature extraction would be hard to notice and would highly impact the success of the project.

Our approach relies on automatically learning the embeddings through an autoencoder instead. An autoencoder consists of an encoder and a decoder, which are trained together to learn efficient representations of input data. This process creates an information bottleneck in the network (the embedding in our application), forcing the encoder to learn a compact representation of the input so that the decoder can reconstruct it with minimum loss. This process essentially embeds the original high-dimensional features (1386 dimensions) to a lower-dimensional feature vector (350 dimensions). The lower-dimensional feature vector retains only the most essential and effective features, so that the input computation vector can be reconstructed faithfully. Consequently, when this learned embedding is used to train the performance model, the model can more efficiently utilize these features for downstream performance prediction. Additionally, the use of these effective features helps mitigate the risk of overfitting, potentially reducing the need for generating more labeled data to achieve robust model performance.

\subsection{Training the Performance Model}

Once the autoencoder is trained, we use its encoder part to embed all the computation vectors before they are fed to the performance model for training. In other words, the embedding by the pre-trained encoder serves as an upstream task, while the recursive embedding of the AST-structured performance model is the downstream task that predicts speedup. 

At the beginning of training, all the weights in the pre-trained encoder are frozen. This is because we do not want the performance model to alter the learned weights in these pre-trained layers, causing catastrophic forgetting~\cite{French_1999}. Instead, the rest of the model should learn how to map the learned embeddings to speedup. However, reconstructing computation vectors from embeddings and predicting speedup are essentially different tasks. After the loss of the performance model (with respect to speedup prediction) stops to decrease for a specified number of epoch, the weights in the pre-trained layers are unfrozen, allowing the weights to be updated by backpropagation. The learning rate for these pre-trained layers is set much smaller ($\times 0.2 $) than that used in the other parts of the performance model to avoid catastrophic forgetting \cite{French_1999}.

\section{Evaluation} \label{sec:evaluation}
To assess the efficacy of our proposed approach, we conduct several experiments. First, we compare the accuracy of the Tiramisu performance model trained with and without a pre-trained encoder (Sec. \ref{subsection:model-accuracy}). Additionally, we investigate whether the observed accuracy improvement under limited data results from our proposed pre-training, or simply because the new model (encoder + performance model) has a more complex architecture. This is achieved by comparing two performance models, both equipped with our encoder, one initialized with pre-trained weights and the other with random weights (Sec. \ref{subsection:stronger-encoder-not-enough}). We further evaluate the impact of pre-training on the quality of code optimizations found by the autoscheduler (i.e., whether pre-training impacts the speedups obtained by Tiramisu's autoscheduler) in Sec. \ref{subsection:effects-on-autoscheduler}. To address concerns about potential slowdowns in autoscheduling (due to the use of an encoder which makes the end-to-end model more complex), we measure the time taken by Tiramisu's autoscheduler when using our proposed method, ensuring minimal impact on efficiency (Sec. \ref{subsection:search-time}). We also include an ablation study on the pre-training network in Sec. \ref{subsection:ablation-study} and discuss the exploration of design choices for our proposed approach that we considered early in the project (Sec.~\ref{subsection:design-choices}).

\paragraph{Machine characteristics} 
We performed all the evaluations on a node with a 28-core Intel(R) Xeon(R) CPU E5-2680 v4 @ 2.40GHz, $4$ GB of RAM per core. The OS installed on the node is CentOS Linux version $8$.

\paragraph{Notations}
In the rest of the paper, we refer to the original performance model from \citet{Merouani_2024} as the \textbf{\texttt{ORIGINAL}} model. We refer to the model that uses a pre-trained encoder (our proposed approach) as \textbf{\texttt{OUR}} model or simply \textbf{\texttt{OURS}}. In some of the evaluations, we train the models on only a subset of the data. For example, we train the \texttt{ORIGINAL} model on $10\%$ of the full training dataset. For simplicity, we call this model \texttt{ORIGINAL-0.1-DATA}, while \texttt{OURS} trained with $10\%$ of the full training datasets is called \texttt{OURS-0.1-DATA}, and the same rule applies to other data sizes. When we train a model on the full dataset we add the suffix \texttt{-FULL-DATA}. For example, we would use (\texttt{ORIGINAL-FULL-DATA}) to refer to the \texttt{ORIGINAL} model trained on the full dataset.

\paragraph{Datasets}
We acquired the dataset for training and testing the performance model from \citet{Merouani_2024}, which contains around $26$ millions data points. Each data point is the triplet $\langle$ program, transformation sequence, execution time $\rangle$. In their latest paper \cite{Merouani_2024}, they use a total of $29$ millions datapoints to train their performance model, but the dataset we used in this paper is a dataset that we obtained at an earlier time when they only had $26$ millions data points. We split the dataset acquired from \citet{Merouani_2024} into a training set (18 M), validation set (3.6 M), and test set (3.6 M).

\begin{figure}[t]
    \centering
    \includegraphics[width=0.47\textwidth]{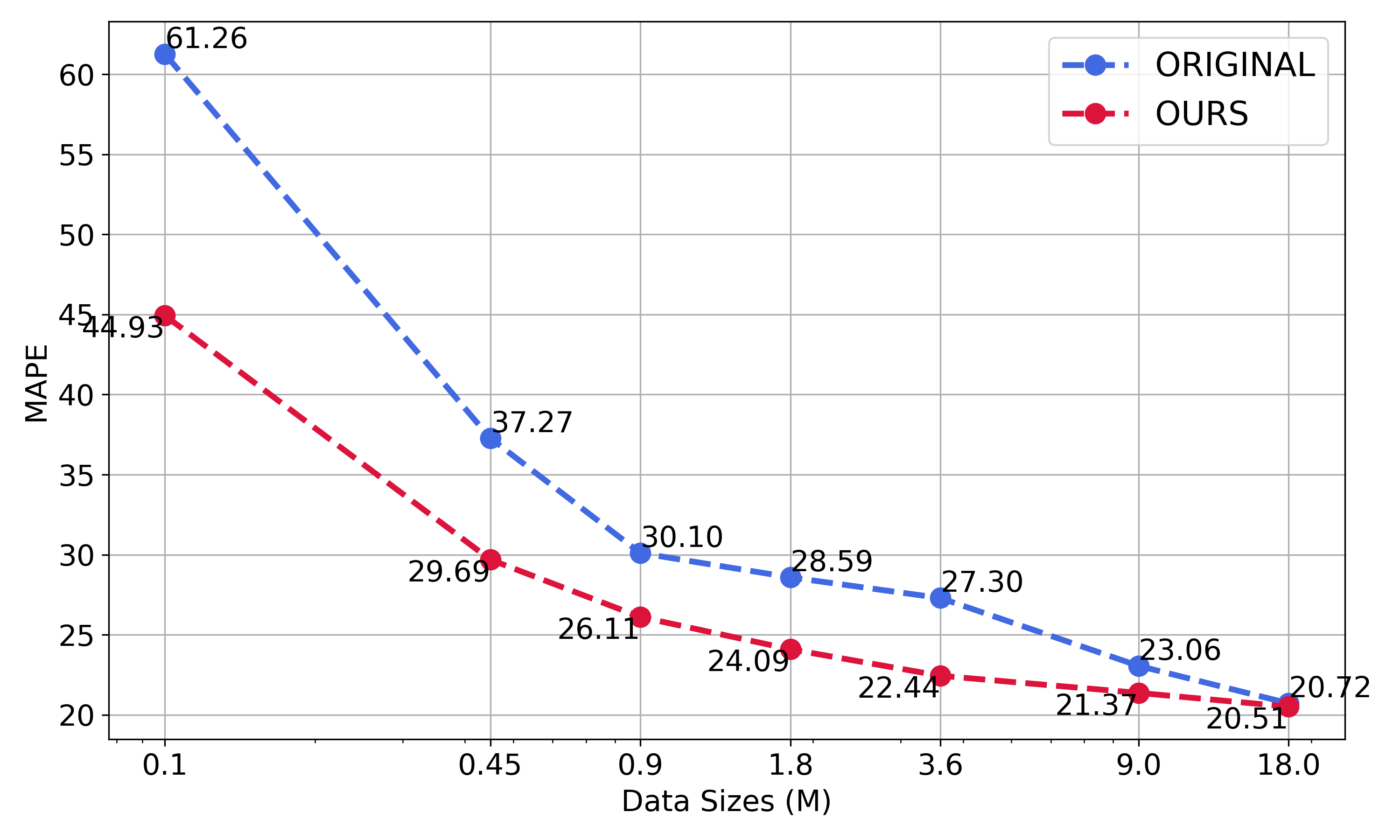}
    \caption{MAPE achieved after training the performance model with (\texttt{OURS}) and without (\texttt{ORIGINAL}) pre-trained autoencoder on different datasize}
    \label{mape}
\end{figure}

\subsection{Performance Model Accuracy}
\label{subsection:model-accuracy}
We first train our proposed autoencoder using the dataset described in Section 5.2, which contains vector representations of code statements and optimizations. The autoencoder is trained to minimize reconstruction loss, resulting in a final pre-trained model with a MSE of $0.0027$. The encoder part of this pre-trained autoencoder is then incorporated into the performance model to generate embeddings and is fine-tuned as outlined in Section \ref{methodology} during the performance model training.

The metric that we use as the loss function and to evaluate the accuracy of the two models after training is the \textbf{MAPE}, which is the Mean of the Absolute Percentage Errors of predictions. The MAPE is calculated as: 
$$MAPE = \frac{1}{N} \sum_{t = 1}^{N}\frac{|A_t - P_t|}{A_t}, $$
where $A_t$ is the measured speedup and $P_t$ is the predicted speedup at the data point $t$. 

When trained on the full (18 M) datapoints, the Mean Absolute Percentage Error (MAPE) achieved by the \texttt{ORIGINAL} model on the test set is $20.72\%$. In order to simulate the situation when training data is limited, we randomly sample from the training dataset to create smaller datasets, containing 9 millions $(50\%)$, 3.6 millions $(20\%)$, 1.8 millions $(10\%)$, 0.9 millions $(5\%)$,  0.45 millions $(2.5\%)$, and 0.1 millions $(0.5\%)$ datapoints respectively. We sample the dataset three times randomly for each one of the previous sizes, and train the \texttt{ORIGINAL} model and \texttt{OURS} on them. For each data size, we report the average MAPE for the three trainings.

Figure \ref{mape} shows the evaluation results for each data size on the test set, averaged over three samples. We note that the MAPE results across the three samples for each data size are highly consistent, with differences of less than $0.3\%$. As we can see from the plot, \texttt{OUR} model achieves much higher accuracy compared to the \texttt{ORIGINAL} model when the data size is less than $9$ millions. In particular, when the dataset has only $0.45$ millions datapoints, which is $2.5\%$ of the full training data, our proposed approach outperforms the \texttt{ORIGINAL} model by $7.58\%$. This shows that our pre-training approach allows the model to have a lower MAPE when there is less  data.

\begin{figure}[h]
\centering
\includegraphics[width=0.3\textwidth]{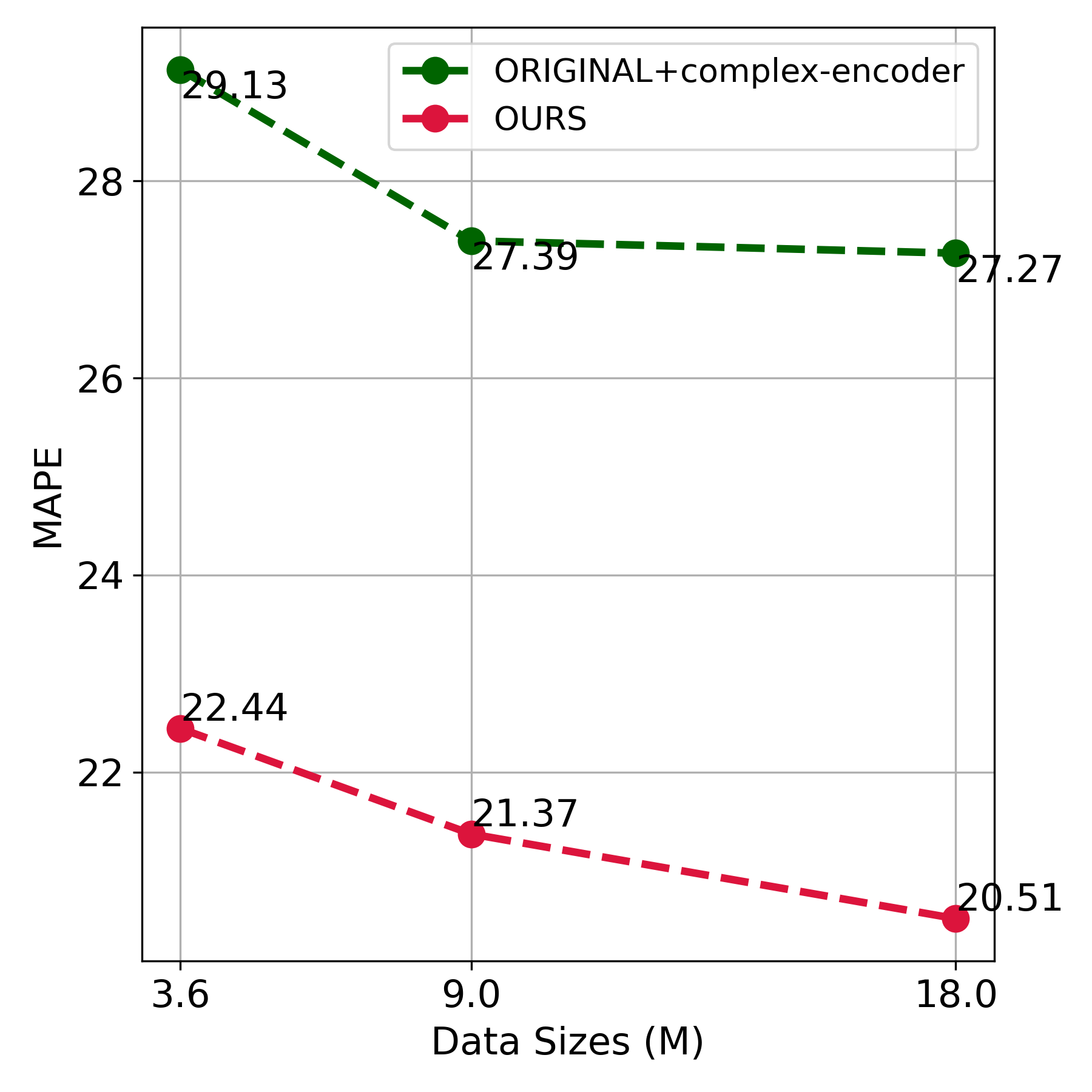}
\caption{MAPE achieved after training two performance models, both utilizing our encoder architecture, but one with pre-trained weights (\texttt{pretrained}) and the other with randomly initialized weights (\texttt{complex-embed}). }
\label{fig:complex_embed}
\shrinkafter
\end{figure}

\begin{figure*}[h]
\centering
\includegraphics[width=0.7\textwidth]{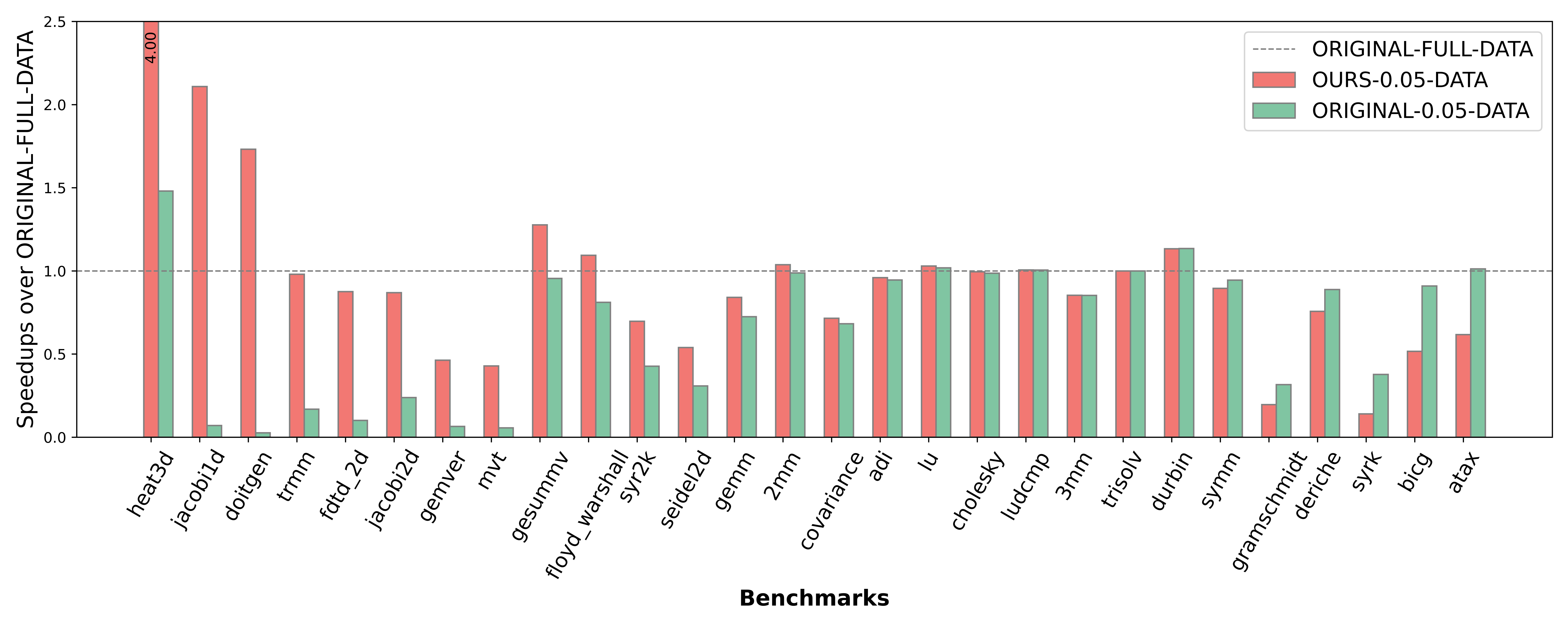}
\includegraphics[width=0.7\textwidth]{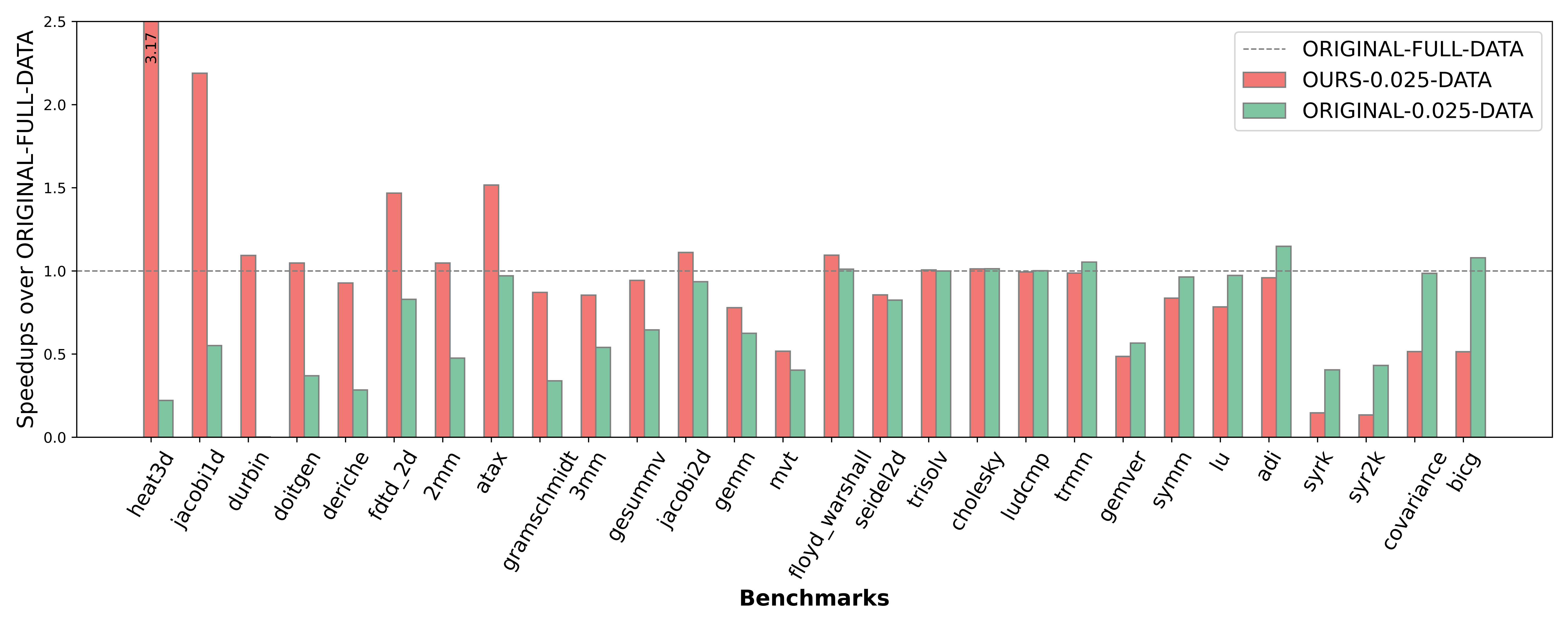}
\includegraphics[width=0.7\textwidth]{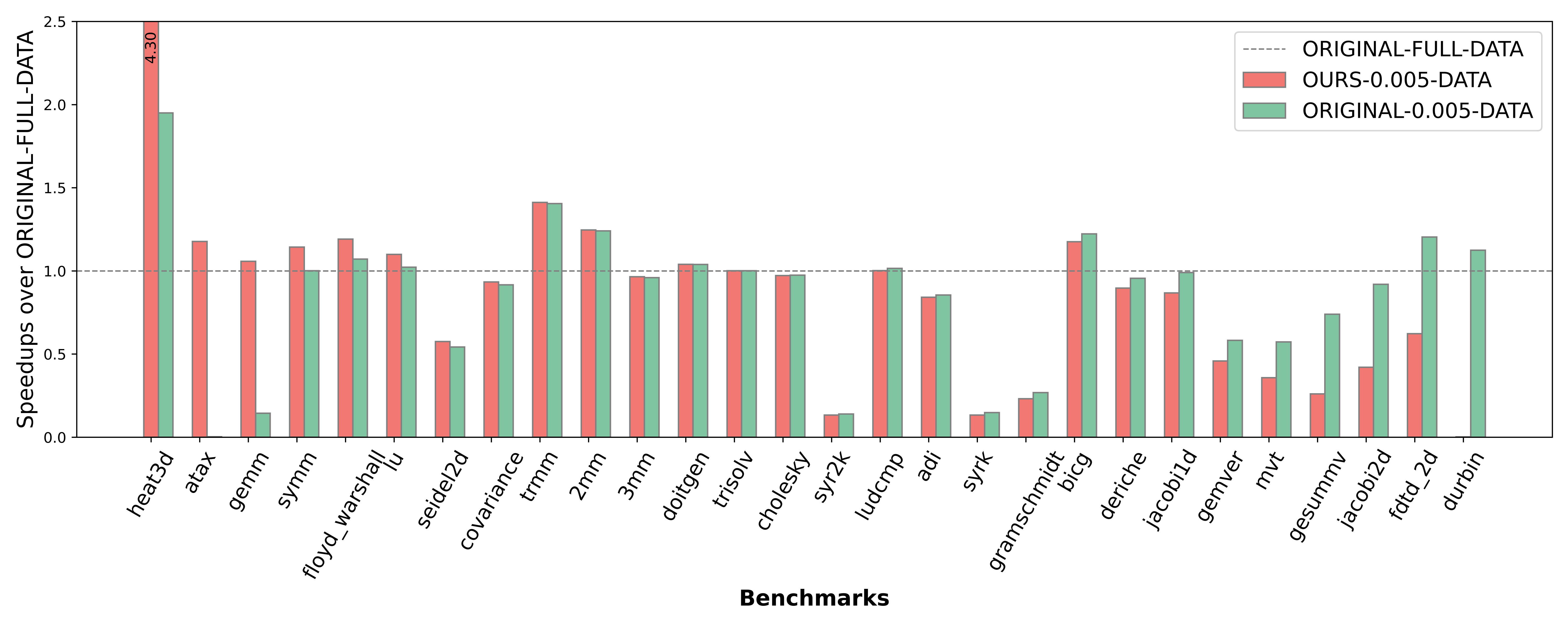}
\caption{Speedups achieved by the Tiramisu's autoscheduler using the \texttt{ORIGINAL} model and \texttt{OUR} model trained on different data sizes ($10\%, 5\%, 1\%$, top to bottom) relative to Tiramisu's autoscheduler with \texttt{ORIGINAL} model trained on the full dataset (\texttt{ORIGINAL-FULL-DATA}) on $28$ benchmarks from PolyBench. }
\label{fig:speedups}
\end{figure*}

\subsection{A Stronger Encoder Is Not Enough}
\label{subsection:stronger-encoder-not-enough}
As discussed in Section \ref{subsection:autoencoder-pretraining}, the proposed encoder architecture exhibits greater complexity and depth compared to the computation embedding layer (found at the leaf of the AST-structured model) in the original performance model. As both our proposed encoder and the computation embedding layers serve as feature extractors from the input tensors, it is important to question whether the observed improvement in accuracy under limited data stems solely from the architectural complexity or from the proposed pre-training scheme. To validate this, we train two performance models, both equipped with our proposed encoder to embed input statements. One model is loaded with pre-trained weights, while the other initializes weights randomly. Figure \ref{fig:complex_embed} illustrates the MAPE achieved after training these models on varying dataset sizes. Notably, without pre-trained weights, the performance of the model equipped with our complex encoder significantly deteriorates, even performing worse than the \texttt{ORIGINAL} model. This demonstrates that the observed improvement in accuracy under limited data stems mainly from the proposed pre-training scheme and not from the architectural complexity after adding the encoder.

\subsection{Effects on the Autoscheduler}
\label{subsection:effects-on-autoscheduler}

Since the performance model is, after all, used by the autoscheduler to evaluate optimizations during the search, we proceed to evaluate how the improvement in accuracy (on a limited size of data) impacts the speedups of code optimized by the autosheduler. We evaluate the performance of the autoscheduler with different performance models on the PolyBnech benchmark suite~\cite{Pouchet_2021}, the same used to evaluate the Tiramisu autoscheduler in LOOPer's paper~\cite{Merouani_2024}. Polybench consists of benchmarks extracted from various computing areas such as linear algebra, stencils, physics simulation, etc.
We use $28$ out of the $30$ benchmarks in PolyBench version 4.2.1, as the $2$ benchmarks are not yet supported by the Tiramisu autoscheduler version we acquired from \citet{Merouani_2024}, at the time these experiments were done.
We compare the speedups achieved by the same autoschedulers, but one equipped with the \texttt{ORIGINAL} model trained with datasets of sizes 0.9 M $(5\%)$, 0.45 M $(2.5\%)$, 0.1 M $(0.5\%)$, while the other with \texttt{OUR} model also trained with the same datasets.
For each benchmark, we use three different representative sizes for the input data as defined by PolyBench (SMALL, MEDIUM, LARGE), and report the geometric mean of the speedups obtained on all three sizes for presentation clarity and simplicity. 

Figure~\ref{fig:speedups} shows the speedups of autoschedulers using the \texttt{ORIGINAL} model and \texttt{OURS} trained on the three smallest data sizes we have ($5\%, 2.5\%, 0.5\%$, top to bottom) relative to Tiramisu's autoscheduler with the \texttt{ORIGINAL} model trained on the full dataset (\texttt{ORIGINAL-FULL-DATA}). For each plot in Fig. \ref{fig:speedups}, benchmarks are sorted by the difference of the speedups between the two models. In particular, \texttt{OURS-0.05-DATA} outperforms \texttt{ORIGINAL-0.05-DATA} on $21$ out of $28$ benchmarks and achieves a geometric mean of $1.83 \times$ over it. Table~\ref{table:speedup-statistics} compares the geometric mean speedup obtained using \texttt{OUR} model and that obtained using the \texttt{ORIGINAL} model (\texttt{OUR}/\texttt{ORIGINAL}). Values above 1 indicate that speedups obtained using \texttt{OUR} model are higher on average compared to those obtained using the \texttt{ORIGINAL} model. We can see in the table that when the performance model is trained with $5\%$ and $2.5\%$ of the dataset \texttt{OUR} model outperforms the \texttt{ORIGINAL} model on most benchmarks. When the performance models are trained with even smaller dataset ($0.5\%$), \texttt{OUR} model performs similarly to \texttt{ORIGINAL}. This indicates that below certain threshold, the MAPE achieved by both models when trained with such less data are too high $(> 40\%)$, such that they both cannot help autoschedulers find good optimizations. Above this data size threshold, the results show that the improved accuracy of the performance model trained with less data enhances the speedup performance of the autoscheduler.

We observe that, on certain benchmarks, a partially trained model can outperform a fully trained one. This outcome likely stems from the complex combinatorial nature of the search problem, where the autoscheduler uses the model’s continuous predictions to navigate a vast space of code transformations. In some cases, minor inaccuracies in a partially trained model may lead the search toward alternative paths that yield better solutions, provided these deviations do not significantly misguide the search. Enhancing the robustness of the search heuristic, however, falls outside the scope of this work.

\begin{table}[h]
\small
\centering
\begin{tabular}{|r|c|c|}
\hline
 & \begin{tabular}{@{}c@{}} $\#$ Benchmarks \\ \texttt{OURS} $>$ \texttt{ORIGINAL} \end{tabular} 
 & \begin{tabular}{@{}c@{}} Speedup ratio \\ (\texttt{OURS} $/$ \texttt{ORIGINAL}) \end{tabular} \\
\hline
0.05-DATA &  21 & $1.89 \times$ \\
\hline
0.025-DATA & 17 & $1.58 \times$ \\
\hline
0.005-DATA &  12 & $0.97 \times$ \\
\hline
\end{tabular}
\caption{The number of benchmarks (out of 28) on which \texttt{OUR} model outperforms the \texttt{ORIGINAL} model and the ratio of the geometric means of the speedups they achieved over benchmarks.}
\label{table:speedup-statistics}
\shrinkafter
\end{table}

\subsection{Effects of the Pre-trained Encoder on the Search Time}
\label{subsection:search-time}
As we are proposing to add an encoder to the performance model, it is important to evaluate how it affects the search time of the autoscheduler. Since the new performance model with the encoder might be significantly slower than the original model without an encoder. Our measurement shows that on average, the search time taken by the autoscheduler with the pre-trained encoder is only $1.05 \times$ slower than that of the autoscheduler that uses the baseline model. We believe that this difference is small in comparison with the benefits obtained when using the pre-trained encoder.

\subsection{Ablation Study for the Pre-training Network} % Use a table
\label{subsection:ablation-study}
Early in our project, we explored various alternatives for the autoencoder networks. Initially, we used the same architecture as the computation embedding layer (Fig. \ref{fig:performance_model_dnn}) in LOOPer as an encoder. We also experimented with a simple MLP as an encoder. Table \ref{table:ablation-study} shows the mean square error achieved by different encoder architectures during pre-training, and MAPE achieved when the performance model equipped with these pre-trained encoder is trained on 1.8 M (0.1-DATA) dataset. We see that both of those two alternative architectures resulted in slightly worse performance in the pre-training (reconstruction) task and the downstream performance modeling task. 

\begin{table}[]
\small
\centering
\begin{tabular}{|c|c|c|c|}
\hline
Encoder architecture & \texttt{OURS} & MLP & LOOPer's Comp \\ 
                     &               &     & Embed Layer \\
\hline
MSE &  \textbf{0.0027} & 0.0055 & 0.0035 \\
\hline
MAPE (0.1-DATA) & \textbf{24.09} & 26.13 & 25.50 \\
\hline
\end{tabular}
\caption{Mean square error achieved by different encoder architecture during pre-training, and MAPE achieved when the performance model equiped with these pre-trained encoder is trained on 1.8 M (0.1-DATA) dataset. }
\label{table:ablation-study}
\shrinkafter
\end{table}

\subsection{Exploring Other Design Choices}
\label{subsection:design-choices}

\paragraph{Hand-engineered features} Our original motivation was to reduce the amount of data needed to train the performance model used in Tiramisu. Before experimenting with the idea of using pre-training, we attempted to hand-engineer features from code to improve performance modeling efficiency. Features we have experimented to include in the computation vector include but not limited to:
\begin{itemize}
    \item Memory access strides.
    \item Size of data accessed in each buffer.
    \item Polyhedral schedule matrix to represent loop transformations.
\end{itemize}

However, none of these attempts was successful. Using an autoencoder to automatically extract high-quality features proved to be the most effective approach.

% Other design choices
\paragraph{LLVM-IR-Based pre-training} As we discussed in Sec. \ref{subsection:learn-code}, many related projects extract the pre-trained embeddings from the LLVM IR (a low level IR). However, using LLVM IR-based embeddings in a search-based compiler is costly. This is because, for every code transformation explored by the search algorithm, the compiler must apply the transformation and compile the transformed code to produce its corresponding LLVM IR and generate the embedding from that IR. For instance, with the PolyBench programs used in Sec. \ref{subsection:effects-on-autoscheduler}, the average time and median time required to apply a sequence of transformations and compile the optimized code in Tiramisu is $1885.9$ milliseconds and $1746$ milliseconds respectively. In contrast, the average inference time of the performance model in Tiramisu is only $32$ milliseconds. Depending on the configuration of the autoscheduler in Tiramisu, such an expensive compilation down to LLVM IR is repeated many times, ranging from $50$ to a few thousands. Therefore, compiling code down to LLVM IR would be expensive. To maintain search efficiency, search-based compilers predict the expected performance when optimizations are applied to a given code without compilation to LLVM-IR, by directly extracting a representation from the source code (or a slightly optimized version of the source code) and feeding it to the performance model. Because of this reason, pre-training methods developed for LLVM-IR are not well suited for search-based compilers.

\section{Conclusion}
We developed an autoencoder-based pre-training scheme to alleviate the data requirements for training performance models used in autoschedulers. Our approach involves pre-training an autoencoder using randomly generated programs and utilizing its encoder part to embed program statements. We demonstrate that our pre-training scheme significantly enhances the accuracy of speedup prediction in the performance model of Tiramisu, particularly when trained on smaller datasets ($< 9$ million data points). Moreover, this improved accuracy reflects on the higher speedups achieved by the autoscheduler that is trained with a pre-trained encoder on a small dataset compared to autoscheduler trained on the same data but without our encoder. Notably, when the Tiramisu performance model was trained on only 0.45 million data points, using our proposed approach, it outperformed the original model by $7.58\%$ MAPE, and an autoscheuler using it achieves speedups $1.58\times$ higher than an autoscheduler trained on the same dataset but without our encoder. Our proposed approach allows the training of performance models with less data, opening the door for a wider adoption of performance models in compilers.

\section{Acknowledgment}
This research has been partly supported by the Center for Artificial Intelligence and Robotics (CAIR) at New York University Abu Dhabi, funded by Tamkeen under the NYUAD Research Institute Award CG010. The research was carried out on the High-Performance Computing resources at New York University Abu Dhabi.

%% Acknowledgments
%\begin{acks}                            %% acks environment is optional
                                        %% contents suppressed with 'anonymous'
  %% Commands \grantsponsor{<sponsorID>}{<name>}{<url>} and
  %% \grantnum[<url>]{<sponsorID>}{<number>} should be used to
  %% acknowledge financial support and will be used by metadata
  %% extraction tools.
  %This material is based upon work supported by the
  %\grantsponsor{GS100000001}{National Science
  %  Foundation}{http://dx.doi.org/10.13039/100000001} under Grant
  %No.~\grantnum{GS100000001}{nnnnnnn} and Grant
  %No.~\grantnum{GS100000001}{mmmmmmm}.  Any opinions, findings, %and
  %conclusions or recommendations expressed in this material are %those
  %of the author and do not necessarily reflect the views of the
  %National Science Foundation.
%\end{acks}

%% Bibliography
\clearpage
\bibliography{main}

%% Appendix
%\appendix
%\section{Appendix}

%Text of appendix \ldots

\end{document}